\documentclass[aps,showpacs]{revtex4}
\usepackage{amsbsy}
\usepackage{graphicx,color}
\usepackage{amsfonts}
\usepackage{amsmath}
\usepackage{amssymb}
\usepackage{amsthm,bm}
\usepackage{dsfont}

\newcommand{\up}{\uparrow}
\newcommand{\down}{\downarrow}
\newcommand{\nn}{\nonumber}

\begin{document}
\title{Trion ground state energy: simple results}
\author{R. Combescot}
\affiliation{Laboratoire de Physique, Ecole Normale Sup\'erieure, PSL Universit\'e, 
Sorbonne Universit\'e, Paris Diderot Universit\'e, CNRS, 24 rue Lhomond, F-75005 Paris,
France.}
\date{Received \today}
\pacs{03.65.-w , 31.15.-p , 71.35.-y, 71.35.Pq}

\begin{abstract}
We investigate the trion binding energy in a three-dimensional semiconductor, with bare Coulomb interaction between
charges, and effective mass approximation for the electron and hole dispersion relations. This is done by making use
of a previously proposed exact method for the three-body problem. The calculations cover the complete
range of electron-to-hole mass ratio. We find a perfect agreement with existing variational calculations. Investigating
the small and large mass ratio regimes, we build a three parameters interpolating formula for the trion binding energy
$E_b(r)$ in terms of the exciton binding energy, where $r$ is the electron to exciton mass ratio. This formula 
$E_b(r)=0.71347-0.11527 \,r -0.18580\, \sqrt{1-r}\,$, in atomic units, is in full agreement, within our precision, with our numerical 
results over the complete range of mass ratio.
\end{abstract}
\maketitle

\section{INTRODUCTION}

Optical properties of semiconductors, mostly in quantum wells and in reduced dimensions, are to a large extent
controlled by their few-body low energy excitations \cite{kezerev}. 
Most important is the exciton, but trions and bi-excitons  are also
quite relevant. Following the prediction of its existence \cite{lampert} and its observation \cite{kheng,huard},
the trion (or 'charged exciton') has been indeed recognized as an important ingredient in understanding optical properties.
In quantum wells the trion may arise, for example, from electrons forming
bound states with photo-created excitons. The extra electron may come from intentional doping, but also from
impurities. In addition to quantum wells, the trion is also important
in understanding the optical properties of quite interesting materials which have been the subject of a number of recent
investigations, such as nanowires \cite{kezma}, carbone nanotubes \cite{carbnano} and transition-metal dichalcogenides
 \cite{chern,berke,sidler,efmac,keztsi}.

It is naturally quite important to know the binding energy of all the excitations in order to identify them 
properly, and ultimately manipulate the optical properties. While in many cases this is achieved for the exciton,
the case of the trion is more difficult. Indeed if there are not enough of them, due for example to low doping,
the trion line is too weak to be seen. On the other hand in the case of high doping, the trion linewidth gets large 
which makes it difficult to identify. Moreover in this case the presence of an electronic Fermi sea makes the
trion problem even more difficult, due in particular to Pauli blocking effects \cite{sycmon,sccepl,efmac}.
The matter is complicated enough to make controversial the identification of the trion line \cite{comtri}.
Since the identification of the trion line is not obvious, precise calculations of its position are clearly of interest
to provide an additional ingredient for the identification.

In a recent paper \cite{rcCoul} we have proposed a general, exact and efficient method for handling the three body problem,
with the aim to apply it to the trion \cite{fad}. The essential idea of this method is to make full use of the solution
of the two-body problem, corresponding to the exciton, to solve the trion problem. The solution of the two-body
problem comes in practice through the knowledge of the corresponding T-matrix.
This approach is of general interest since the trion problem is not an easy one, and
it is often simplified in order to achieve an effective solution.
Our method leads to solve a simple integral equation, a task in principle easier than what has to be done with more sophisticated methods.
This is at least an alternative method of calculation for trion properties, which can complement other approaches.

Let us stress that the interest of our method is not restricted to the case where the electron-hole Hamiltonian corresponds to
the effective mass approximation, which we will use here, nor to the three-dimensional case we will consider. 
Indeed it can be applied in principle
to any situation, for example quantum wells, 2D systems \cite{chern,berke,sidler,efmac,keztsi} or embedded quasi-2D systems \cite{tuan}, 
where the electron and hole dynamics is more complicated. Similarly it is not restricted to the bare Coulomb interaction
and there is in principle no problem to take screening into account and use for example the Keldysh potential
\cite{keld}. All this makes the calculation
of the exciton properties more involved. However the point is that, as soon as the exciton is known through
its T-matrix, the trion energy can be calculated within exactly the same framework. This insures that the
trion properties are obtained within exactly the same theoretical description as the exciton properties,
avoiding any incoherence which might arise from practical problems when the two objects are handled independently.
Another quite important interest of the method is that it provides, at the same time as the binding energy, the trion
wave function, although we will not make use of it in the present paper. Finally its formalism is well suited for
extensions to many-body problems, such as the trion in the presence of the electronic Fermi sea.

We have applied \cite{rcCoul} our method to the specific case of the Helium atom ground state to check its validity and efficiency.
In this specific case, since we have to deal with the Coulomb interaction, we have used the very convenient expression 
for the T-matrix obtained by Schwinger \cite{schwing}, which is almost analytical, even if it is somewhat singular.
This check on the Helium case has been completely satisfactory. However it must be mentioned that,
in retrospect, this was dealing with a fairly easy situation. Indeed in the Helium case, the repulsive
interaction between the two electrons is fairly weak compared to the attractive interaction between the
nucleus and the two electrons. Actually a perturbative calculation gives already an approximate result
which is not so bad. Naturally our calculation has been an exact one, dealing with the full three-body
problem. But the conditions were good and the numerical implementation could go quite smoothly.

Here we want to use this method to actually calculate the trion binding energy.
As a first case, we will deal with the situation which is formally the simplest one, namely the three-dimensional case
within the effective mass approximation for electron and hole. The interaction is the bare Coulomb interaction
so we can again use Schwinger's T-matrix. This case has already been investigated \cite{usu,sycmon}, 
so we can compare our results to existing work. 
Our aim is to investigate the effect of the electron-hole mass ratio,
and see how important it is. Despite its apparent simplicity this case turns out to be a fairly stringent test of
our method, because the trion is quite weakly bound with respect to the exciton. Very roughly speaking
the trion and the exciton are almost degenerate. Clearly such a situation leads to complications, which are
probably present whatever the method used to solve the problem. As we will see this physical situation
makes indeed our calculations in practice somewhat more complicated that in the Helium case.

In most compounds the electronic effective mass $m_e$ is lighter or of the order of the hole mass $m_h$,
so in the standard negatively charged trion, with two electrons and one hole corresponding to the 
$(eeh)$ or $(X^-)$ situation, the mass ratio $m_e/m_h$ is smaller than unity or of the same order. However there is
also the possibility of a positively charged trion, with two holes and one electron corresponding to the 
$(ehh)$ or $(X^+)$ situation. However this second case is formally identical to the first one, provided we exchange
the roles of electrons and holes. Accordingly it corresponds to the case where the mass ratio $m_e/m_h$ is larger 
than unity. Hence dealing with the general situation is equivalent to study the $(eeh)$ trion with a mass ratio having any value.

We will pay particular attention to the limiting domains where this mass ratio is either small or large.
These corresponds to the cases where the trion is physically similar to the negatively charged hydrogen ion $H^-$,
or to the positively charged hydrogen molecule $H_2^+$. It is in these regimes that the dependence of the trion
binding energy on the mass ratio is stronger. In between the binding energy has a minimum,
and the domain where it takes values near this minimum is fairly extended. As a result the trion binding energy
is very weakly dependent on the mass ratio in a wide domain, a fairly surprising physical feature. Hence an understanding
of the binding energy in the limiting domains of small and large mass ratio leads to a better grasp of the
physical situation in between. Elaborating quantitatively on this principle, we have in addition built a simple interpolation 
formula which describes properly the behaviour of the binding energy in these two limiting regimes. Very surprisingly 
this formula happens to be in perfect agreement with our numerical results in the whole range of mass ratio. 
The agreement is so good that, for any practical purpose, there is no need of further numerical calculations to have 
in this physical situation the binding energy for any mass ratio, as if we had an analytical result.

In the next section we summarize our formalism and discuss its implementation. The following section considers
the limiting domains of low and high mass ratio. We give then our results for the binding energy, compare them
to the literature and provide our interpolation formula. The paper closes with a summary.

\section{Formalism}\label{form}

The approach we have used in \cite{rcCoul} is a diagrammatic approach.
It makes full use of the solution of the two-body problem since a basic ingredient is the
T-matrix corresponding to this problem. In the case of the Coulomb interaction this $T_2$ matrix
has a quite simple form found by Schwinger \cite{schwing}. From this knowledge
an integral equation for the scattering amplitude $T_3$ of an electron on the exciton is obtained.
Basically the energy spectrum of the trion, and more specifically its ground state energy is found
from the poles of this scattering amplitude $T_3$. Actually in order to find the informations we are
looking for we only need \cite{rcCoul} to consider the on-the-shell expression for $T_3$.
More specifically we need a symmetrized (corresponding to the electronic singlet expected for the trion ground
state) amplitude $T({\bf p},{\bf p}')$ where ${\bf p}$ and ${\bf p}'$ are the momenta of the $\up$ and $\down$ electron
respectively. Moreover we need an analogous amplitude $S({\bf Q},{\bf q})$ for electron-hole scattering.

We have found \cite{rcCoul} that, for $E > 0$ to be the trion binding energy, $T({\bf p},{\bf p}')$ and $S({\bf Q},{\bf q})$ must
satisfy the following homogeneous linear integral equations:
\begin{eqnarray}\label{eqThom}
T({\bf p},{\bf p}')&=&
-\frac{1}{E+\frac{{\bf p}^2+{\bf p}'^2}{2m_e}+\frac{({\bf p}+{\bf p}')^2}{2m_h}} \\  \nn
&\times &\!\!\sum_{\bf k}\Bigg[T_2\!\left(\!\{-E{-}\frac{{\bf p}'^2}{2m_e},{-}{\bf p}'\};{\bf p}{+}r{\bf p}',{\bf k}{+}r{\bf p}'\right)\;T({\bf p}',{\bf k})
+T_2^e\!\left(\!\{-E{-}\frac{({\bf p}{+}{\bf p}')^2}{2m_h},{\bf p}{+}{\bf p}'\};\frac{{\bf p}{-}{\bf p}'}{2},{\bf k}\right)\;S({\bf p}+{\bf p}',{\bf k})\Bigg]
\end{eqnarray}
\begin{eqnarray}\label{eqShom}
\hspace{-12mm} S({\bf Q},{\bf q})&=&
-\frac{1}{E+\frac{{\bf Q}^2}{2m_h}+\frac{{\bf Q}^2+4{\bf q}^2}{4m_e}}  \\  \nn
&\times &\sum_{\bf k}\;T_2\left(\{-E{-}\frac{(\frac{{\bf Q}}{2}+{\bf q})^2}{2m_e},{-}(\frac{{\bf Q}}{2}+{\bf q})\};
(\frac{{\bf Q}}{2}{-}{\bf q}){+}r(\frac{{\bf Q}}{2}{+}{\bf q}),{\bf k}{+}r(\frac{{\bf Q}}{2}{+}{\bf q})\right)\;T(\frac{{\bf Q}}{2}{+}{\bf q},{\bf k})
+({\bf q}\leftrightarrow\!-{\bf q})
\end{eqnarray}
Here $m_e$ and $m_h$ are the electron and hole mass respectively, and $r=m_e/M$ is the ratio between 
the electron mass and the exciton total mass $M=m_e+m_h$.  
$T_2(P;{\bf k},{\bf k}')=T_2(\omega,{\bf k},{\bf k}')$ is the electron-hole $T_2$ matrix, where
$P = \{\Omega,\textbf{P}\}$ is an energy-momentum four-vector and $\omega=\Omega -\textbf{P}^2/2M $.
Similarly $T_2^e$ is the $T_2$ matrix for the propagation
of an electron pair (and here the total mass is $M'=m_e+m_e=2 m_e$). The specific expressions resulting from the
Schwinger $T_2(\omega,{\bf k},{\bf k}')$ are given below.

It is convenient to rewrite these equations with reduced units, taking $a_0=4\pi \epsilon/(2\mu _{eh} e^2)$ as unit of length,
with $\epsilon $ the medium permittivity and $\mu_{eh} = m_e m_h/(m_e +m_h)$ the reduced exciton mass. Similarly we
take $1/a_0$ as unit wavevector and $1/(2\mu _{eh} a_0^2)$ as energy unit (this is twice the usual atomic unit).
In this way, with $K^2=2\mu _{eh}a_0^2 E$, Eq.(\ref{eqThom}) and Eq.(\ref{eqShom}) become, with all wavevectors
being now in reduced units,
\begin{eqnarray}\label{eqThominfred}
t({\bf p},{\bf p}'){=}{-}\frac{1}{K^2{+}(1{-}r)({\bf p}^2{+}{\bf p}'^2){+}r({\bf p}{+}{\bf p}')^2}
\int \!\!\frac{d{\bf k}}{(2\pi )^3}\bigg[t_2(\kappa ,{\bf p}{+}r{\bf p'},{\bf k}{+}r{\bf p'})\;t({\bf p}',{\bf k})
+t_2^e(\kappa_e,\frac{\bf p_-}{2},{\bf k})\;s({\bf p_+},{\bf k})\bigg]
\end{eqnarray}
\begin{eqnarray}\label{eqShominfred}
\hspace{-17mm} s({\bf Q},{\bf q}){=}{-}\frac{2}{2 K^2{+}(1{+}r){\bf Q}^2{+}4(1{-}r){\bf q}^2}
\int \!\!\frac{d{\bf k}}{(2\pi )^3}\;t_2(\kappa_Q,{\bf Q}_{-}{+}r {\bf Q}_{+},{\bf k}{+}r {\bf Q}_{+})\;t({\bf Q}_+,{\bf k})
+({\bf q}\leftrightarrow\!-{\bf q})
\end{eqnarray}
where we have used the abbreviations ${\bf p}_{\pm}={\bf p}\pm{\bf p}'$, ${\bf Q}_{\pm}={\bf Q}/2\pm{\bf q}$,
$\kappa =\left[K^2+(1-r^2){\bf p}'^2\right]^{1/2}$, $\kappa_Q=\left[K^2+(1-r^2){\bf Q_+}^2\right]^{1/2}$ 
and $\kappa_e=\left[(1-r)K^2/2+(1-r^2){\bf p}_+^2/4\right]^{1/2}$,
and we have for $t_2$ and $t_2^e$ the explicit expressions:
\begin{eqnarray}\label{eqt2t2e}
t_2(\kappa,{\bf q},{\bf q}')=-\frac{4\pi z}{({\bf q}-{\bf q}')^2}\int_{0}^{1}du\,\frac{u^{-\frac{1}{2\kappa}}(1-u^2)}{[u+z(1-u)^2]^2}\hspace{15mm}
t_2^e(\kappa_e,{\bf q},{\bf q}')=\frac{4\pi z_e}{({\bf q}-{\bf q}')^2}\int_{0}^{1}du\,\frac{u^{\frac{1}{4\kappa_e}}(1-u^2)}{[u+z_e(1-u)^2]^2}
\end{eqnarray}
with, in $t_2(\kappa,{\bf q},{\bf q}')$, $z=(\kappa^2+{\bf q}^2)(\kappa^2+{\bf q'}^2)/[4\kappa^2 ({\bf q}-{\bf q}')^2]$, and, in 
$t_2^e(\kappa_e,{\bf q},{\bf q}')$, $z_e=(\kappa_{er}^2+{\bf q}^2)(\kappa_{er}^2+{\bf q'}^2)/[4\kappa_{er}^2 ({\bf q}-{\bf q}')^2]$ with $\kappa_{er} =\kappa_e/(1-r)$. One notes that the first term in the second equation for $s({\bf Q},{\bf q})$ is merely obtained
from the first term in the first equation for $t({\bf p},{\bf p}')$ by the substitutions $ {\bf p} \to {\bf Q}_{-}$ and ${\bf p'} \to {\bf Q}_{+}$.

The practical handling of these equations follow the same lines as those used in \cite{rcCoul}. However in practice
the numerical work for the trion has turned out to be somewhat more complicated than for the Helium case. A first problem comes
from the iteration procedure we have followed to solve the coupled integral equations Eq.(\ref{eqThominfred})
and Eq.(\ref{eqShominfred}). Indeed these equations can be understood as meaning that the linear operator
corresponding to the right-hand side of these equations has an eigenvalue equal to 1, and, for $E$ to be the
ground state energy, it is easily seen \cite{rcCoul} that this is the largest positive eigenvalue which must be equal to 1.
In principle this largest eigenvalue can be found by merely iterating \cite{rcCoul} the operator corresponding to the right-hand side.

However this procedure converges efficiently only if the spectrum of this operator is well behaved.
In looking for the trion ground state we have seen the appearance of complex eigenvalues with large modulus, which spoil
totally the iteration. Fortunately this has been completely solved by merely by carrying the explicit expression for
$s({\bf Q},{\bf q})$ given by Eq.(\ref{eqShominfred}) into Eq.(\ref{eqThominfred}). This is no problem numerically and this
happens to eliminate our spurious complex eigenvalues.

Another problem we have found is the appearance of negative eigenvalues with large absolute values. However one can
in principle eliminate this kind of problem by shifting around the spectrum by linear transformations. If an operator $A$
has a spectrum included in $[-a,b]$ (with $a,b >0$), the operator $(A+a \mathds{1})/(a+b)$ has its spectrum included
in $[0,1]$ so the largest eigenvalue can be obtained by iteration. Naturally this slows down the convergence of the
iteration, but we have found that this solves all our problems of this kind.

A more systematic problem is linked to the fact that the trion is weakly bound with respect to the exciton.
If this relative binding was zero, with the trion binding equal to the exciton one, we would find cases where $T_2$
would diverge. Specifically the exciton binding energy corresponds to $K=1/2$, so for ${\bf p'}=0$ we have 
$\kappa=1/2$ and the integral for $t_2(\kappa,{\bf q},{\bf q}')$ in Eq.(\ref{eqt2t2e}) diverges. Naturally the total
trion binding energy is slightly larger that the exciton binding energy, and we are not in this singular situation.
Nevertheless we are near this singularity, and in practice this means that the corresponding integrals, while
not being infinite, are much more sensitive to the values of the various parameters than in a standard situation. 
This makes automatically a precise numerical calculation more difficult to implement.

In particular, to avoid the $u$ region responsible for the nearly divergent behaviour,
we have found useful to avoid the low $u$ region (for example the $[0,0.5]$ domain)
in the integral for $t_2(\kappa,{\bf q},{\bf q}')$ 
in Eq.(\ref{eqt2t2e}) by following Schwinger \cite{schwing} and transforming it into a contour integral on a circle going
around the origin $u=0$. Writing explicitly this contour integral, and skiping details, this amounts finally to write
\begin{eqnarray}\label{eqcontour}
 \int_{0}^{1/2} \!du \,f(u)= -\frac{1}{\sin(\frac{\pi }{2\kappa })} \;
 {\mathrm Re}\left[ e^{i\frac{\pi }{2\kappa }}  \int_{0}^{\pi } \!\!d\theta \; u\,f(u) \right]
\end{eqnarray}
with $u = e^{i\theta}/2$ in the right-hand side integral. Here $f(u)$ is the integrand in the integral
for $t_2(\kappa,{\bf q},{\bf q}')$ in Eq.(\ref{eqt2t2e}), which behaves as $u^{-1/2\kappa}$ when $u \to 0$.
In the right-hand side the possible singular behaviour for $\kappa \to 1/2$, for example, appears explicitly
in the factor $\sin^{-1}(\pi /(2\kappa))$. Note that one can not extend this procedure to the whole
$[0,1]$ domain because one has to avoid possible contributions coming from poles of $f(u)$.

Finally it is worth noticing that the case $r=1$ is somewhat singular, since divergences appear in some of
the quantities we have defined. This implies a specific treatment to remove these divergences. We have not
proceeded to this treatment in the general case. However it is of interest to consider the particular case
where the two electrons are not interacting, because our integral equations for this simpler problem can be fully solved analytically.
Indeed in this case we can ignore $s({\bf Q},{\bf q})$ and set it to zero, since it comes directly from the
interaction between the two electrons. We are left with Eq.(\ref{eqThominfred}) for $t({\bf p},{\bf p}')$
which becomes for $r=1$
\begin{eqnarray}\label{eqThominfredr1}
t({\bf p},{\bf p}'){=}{-}\frac{1}{K^2{+}({\bf p}{+}{\bf p}')^2}
\int \!\!\frac{d{\bf k}}{(2\pi )^3}\;t_2(K ,{\bf p}{+}{\bf p'},{\bf k}{+}{\bf p'})\;t({\bf p}',{\bf k})
\end{eqnarray}
This equation is clearly compatible with a solution where $t({\bf p},{\bf p}')=\tau({\bf p}+{\bf p}')$ depends
only on the single variable ${\bf p}+{\bf p}'$. This is actually simple to understand physically. The case $r=1$
corresponds to the situation where the two electrons are infinitely heavy and, in the absence of interaction
bewteen them, they interact only with the hole. It is physically fairly clear that the binding of the hole to the
electrons will be the strongest if the two electrons are located at the same place. This is then just the problem
of a single hole in the presence of a doubly charged electron. Hence the binding energy is four times
the exciton binding energy, that is 4 Rydberg=2 Hartree, which corresponds to $K=1$ with our reduced units.
Since we deal only with the hole, it is natural to see only its momentum ${\bf p}_h$ appearing. But we are
in a referential where the total momentum is zero, so that ${\bf p}_h=-({\bf p}+{\bf p'})$, which explains why
only the sum of the electronic momenta ${\bf p}+{\bf p}'$ appears. Finally we may guess that the wave function,
which is \cite{rcCoul} just $t({\bf p},{\bf p}')$ in the present case, is directly related to the excitonic ground state
wave function $1/(1+p_h^2)$. However since we deal with the wave function for the two electrons it is rather 
reasonable to guess that there is one such contribution for each electron so that
\begin{eqnarray}\label{eqwavefr1}
t({\bf p},{\bf p}')= \left[\frac{1}{1+({\bf p}+{\bf p'})^2}\right]^2
\end{eqnarray}
One can insert these answers in Eq.(\ref{eqThominfredr1}) and one can perform analytically all the
resulting integrations to check that they indeed satisfy Eq.(\ref{eqThominfredr1}). The check can also
be easily performed numerically.

\section{Limiting cases}\label{limca}

Before going to our results for the generic situation, it is useful to 
consider the two limiting cases of very heavy and very light hole mass where the mass ratio $r=m_e/(m_e+m_h)$
goes either to $0$ or $1$. The Hamiltonian is
\begin{eqnarray}\label{hamil}
H=\frac{1}{2 m_e}({\bf p}_1^2 +{\bf p}_2^2)+\frac{1}{2 m_h}{\bf p}_3^2 + V({\bf r}_1-{\bf r}_2)-V({\bf r}_1-{\bf r}_3)-V({\bf r}_2-{\bf r}_3)
\end{eqnarray}
where $V({\bf r})=e^2/(4\pi \epsilon r)$ is the Coulomb interaction.

For $r \to 0$ (corresponding to $m_h \gg m_e$) we follow Bethe and Salpeter \cite{betsal}. 
Going to the center of mass referential where ${\bf p}_3=-({\bf p}_1+{\bf p}_2)$, 
the kinetic energy part $H_c$ of the Hamiltonian becomes
\begin{eqnarray}\label{hamilc}
H_c=\frac{1}{2 \mu_{eh}}({\bf p}_1^2 +{\bf p}_2^2)+\frac{{\bf p}_1 \cdot {\bf p}_2}{m_h}
\end{eqnarray}
where $\mu_{eh} =m_e m_h/(m_e+m_h)$ is the exciton reduced mass. In the limit $r \to \infty$ only the first term is present.
Hence the effect of a finite $m_h$ (in addition to the trivial modification of the exciton reduced mass) is to introduce the
last term. This term gives the 'mass-polarization correction', which turns out to be positive for the ground state energy. 
This mass-polarization term
has been studied quite recently in details by Filikhin \emph{et al} \cite{kezer}. For our purpose, for large $m_h$, the
mass-polarization correction can be evaluated with the $m_h = \infty$ wave function, and the corresponding term
in the energy is merely proportional to $1/m_h \propto r$ in this limit. Since it is positive it gives to
the trion binding energy a negative contribution, linear in $r$.

For the other limit $r \to 1$ (corresponding to $m_e \gg m_h$) it is more appropriate to take as new variables the position
of the light mass, i.e. the hole, with respect to the center of mass of the two heavy ones 
$\boldsymbol \rho ={\bf r}_3 - ({\bf r}_1+{\bf r}_2)/2$,
together with half the distance between these two heavy masses ${\bf R}=({\bf r}_2-{\bf r}_1)/2$.
Taking again the center of mass referential, so that ${\bf p}_3=-({\bf p}_1+{\bf p}_2)$, the conjugate momenta are
$\pi _{\bf R}={\bf p}_2-{\bf p}_1$ and $\pi_{\rho }={\bf p}_3$ and the Hamiltonian is
\begin{eqnarray}\label{hamilb}
H=\frac{1}{4 m_e}{\bf \pi }_{\bf R}^2 +\frac{1}{4}\left(\frac{1}{m_e}+\frac{2}{m_h}\right){\bf \pi }_{\rho}^2 + 
V(2 {\bf R})-V(\boldsymbol \rho +{\bf R})-V(\boldsymbol \rho -{\bf R})
\end{eqnarray}

In our limit this Hamiltonian can be handled by a Born-Oppenheimer treatment. In the case $m_e=\infty$,
we have to deal with the problem of the hole in the presence of the attractive interaction by the two electrons,
which are at fixed positions $\pm {\bf R}$. This corresponds to the Hamiltonian Eq.(\ref{hamilb}) without the
electronic kinetic energy term $(1/4 m_e){\bf \pi }_{\bf R}^2$. Let us call $E_{\infty}({\bf R})$ the corresponding
ground state energy (including the repulsive $V(2 {\bf R})$ contribution). Let us call ${\bf R}_0$ the value of
${\bf R}$ for which its minimum is found.
Considering now the case where $m_e$ is quite large, but not infinite, quantum fluctuations produce a departure
of ${\bf R}$ from this minimum position ${\bf R}_0$. We set ${\bf R}={\bf R}_0 + {\bf u}$. For these fluctuations
we can consider $E_{\infty}({\bf R})$ as an effective potential energy. Hence \cite{remark} we have to handle the Hamiltonian
\begin{eqnarray}\label{hamilBorn}
H_{BO}=\frac{1}{4 m_e}{\bf \pi }_{\bf R}^2 + E_{\infty}({\bf R})
\end{eqnarray}
For large $m_e$ the fluctuations are small, so that we may expand $E_{\infty}({\bf R})$ to lowest order around its minimum
$E_{\infty}({\bf R})=E_{\infty}({\bf R}_0)+ a {\bf u}^2$. In this way the Hamiltonian becomes
\begin{eqnarray}\label{hamilBorna}
H_{BO}=\frac{1}{4 m_e}{\bf \pi }_{\bf u}^2 + a {\bf u}^2+ E_{\infty}({\bf R}_0)
\end{eqnarray}
This is the Hamiltonian of a three-dimensional harmonic oscillator, and accordingly the ground state energy is given by
$E=E_{\infty}({\bf R}_0)+(3/2) \hbar (a/m_e)^{1/2}$. In Eq.(\ref{hamilBorna}) we have assumed for the simplicity of the
presentation that $E_{\infty}({\bf R})$ is isotropic around its minimum. Clearly this is not correct in general and we should
rather introduce the components ${\bf u}_{\parallel}$ and ${\bf u}_{\perp}$ of ${\bf u}$ parallel and perpendicular to 
${\bf R}_0$, with corresponding coefficients $a_{\parallel}$ and $a_{\perp}$ for the expansion, so the parallel and
perpendicular harmonic oscillators do not have the same frequencies. But obviously this does not change the final
conclusion that $E-E_{\infty}({\bf R}_0)$ is proportional to $m_e^{-1/2}$. Since in this large $m_e$ situation
$r=m_e/(m_e+m_h) \simeq  1-m_h/m_e$, so $m_e^{-1/2} \propto (1-r)^{1/2}$, and since the binding energy $E_b(r)$ is
essentially the opposite of the energy, we come to the conclusion that, in the vicinity of $r=1$, $E_b(r)$ behaves as
$-(1-r)^{1/2}$.

\section{Results}\label{result}

\begin{table}
\centering
\begin{tabular}{| l | l |} \hline \hline
r    & Binding energy $E_b(r)$ (a.u.)  \\ \hline
0.   &   0.5282 \\ \hline
.05  &  0.5267 \\ \hline
.1    &  0.5256 \\ \hline
.15  &   0.5247 \\ \hline
.2   &   0.5240 \\ \hline
.25  &  0.5235 \\ \hline
.3   &   0.5233 \\ \hline
.35 &   0.5233 \\ \hline
.4   &   0.5235 \\ \hline
.45  &  0.5239 \\ \hline
.5    &   0.5246 \\ \hline
.55  &  0.5256 \\ \hline
.6    &  0.5270 \\ \hline
.65  &   0.5287 \\ \hline
.7    &  0.5310 \\ \hline
.75  &   0.5340 \\ \hline
.8    & 0.5378 \\ \hline
.85  &   0.5430 \\ \hline
.9    &  0.5503 \\ \hline
.95  &   0.5621 \\ \hline
.97  &   0.5696 \\ \hline
.98  &   0.5748 \\ \hline
.99  &   0.5821 \\ \hline \hline

\end{tabular}
\caption{Trion total binding energy $E_b(r)$ in atomic units as a function of the mass ratio 
of the electron to the exciton mass $r=m_e/(m_e+m_h)$. The exciton binding energy is $0.5$.}
\label{tabres}
\end{table}

\begin{figure}
\centering
{\includegraphics[width=\linewidth]{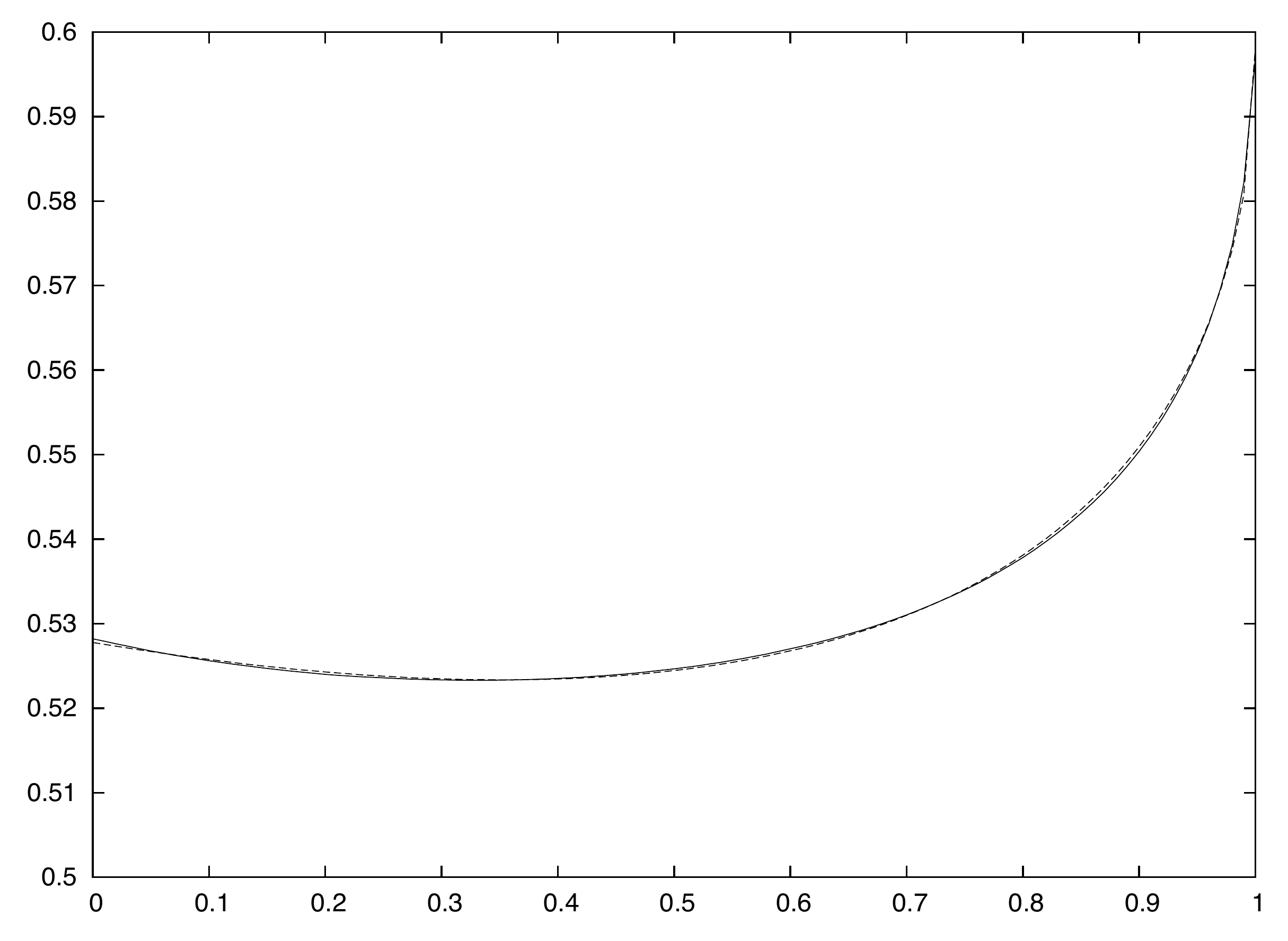}}
\caption{Trion total binding energy $E_b(r)$ in atomic units as a function of the mass ratio of the electron 
to the exciton mass $r=m_e/(m_e+m_h)$. On this scale the exciton binding energy is $0.5$.
Full line: spline interpolation through the results of Table \ref{tabres}. 
Dashed line: fitting formula $E_b(r)=0.71347-0.11527 \,r -0.18580\, \sqrt{1-r}$.}
\label{fig}
\end{figure}

Our numerical results for $E_b(r)$ are given in Table \ref{tabres}. Basically we have made these calculations for 
$r=0.05\,n$ with $n$ going from $0$ to $20$. A finer mesh is unnecessary since the result for any $r$ can be obtained
from our results by an appropriate interpolation without significant loss of precision. This is what we have done in Fig.~\ref{fig}
where we have plotted $E_b(r)$ by making use of a spline interpolation. We have compared our results with those
of Usukura et al \cite{usu} for the mass ratios they have considered. They make use of the stochastic variational method with
correlated Gaussian basis. The agreement is quite fair, and it might very well be
that the small differences are mostly due to the presentation of the results with respect to their precision.

By comparison of our $r=0$ result with much more precise variational calculations \cite{frolov,jcp} which give a binding energy
of $E_b(0)=0.527751$ a.u., we see that we obtain a precision of $10^{-3}$, which somewhat less than the $10^{-4}$ we had in our preceding
Helium calculation, but which is nevertheless quite enough for any practical goal. 
As we have indicated this loss of precision is directly linked to the fact that the difference between the total trion energy
and the exciton energy is small. As a result when we subtract from our total trion energy
the exciton binding energy of 0.5 a.u., we obtain 0.0282 a.u. for the binding energy of the second electron to the exciton, 
which is quite often called the trion binding energy. Our precision is then only about $2.10^{-2}$, 
because this binding is quite small. This is nevertheless suffisant for our purpose.

On the opposite side, for $r=1$, our integral equation becomes singular as we have already mentioned, and accordingly 
we have not done the corresponding numerical calculation. 
On the other hand, as it can be seen in Table \ref{tabres}, we have carried out calculations
up to $r=0.99$, which is quite close to $r=1$. Our corresponding results have to be compared with
variational calculations \cite{koro,li} giving $E_b(1)=0.597139$ a.u. They turn out to be perfectly compatible with
the analytical behaviour $E_b(1)-E_b(r) \propto (1-r)^{1/2}$ we have found in the preceding section \ref{limca}. 
Hence we have incorporated this value for $E_b(1)$ in our results
to obtain Fig.~\ref{fig}, where the full line is obtained from our results in Table \ref{tabres} by a spline interpolation.
One would expect a somewhat degraded precision in close vicinity of $r=1$, which is singular of our calculations.
However we see no clear indication of such a loss at the level of our precision.

We believe that our precision is merely limited by our mesh in the modulus of wavevectors for $t({\bf p},{\bf p}')$: 
we have not been beyond 60 points, because
the computer time increases markedly when one goes to finer mesh (40 points were quite enough in the Helium case). 
One could think of more sophisticated ways to go around
this problem, but there is no real point for such an effort since precision is not our primary purpose. On the other hand, 
with respect to the dependence on the angle between ${\bf p}$ and ${\bf p}'$, we have been near a full convergence 
in this variable by going with Legendre polynomials up to $\ell = 20$,, so we have been able
to extrapolate in $\ell $ to improve our results. However it is worth to note that the need to go to such high values of $\ell$
($\ell =5$ was enough in the Helium case) seems to imply that the wave function is very structured, which deserves clearly
further investigation. This is postponed to a further work.

Considering the dependence of the trion binding energy on the electron-hole mass ratio one has naturally to keep
in mind that the most important contribution comes trivially from the fact that our atomic energy unit is twice the exciton
binding energy, which implies a result proportional to the exciton reduced mass $\mu_{eh} =m_e m_h/(m_e+m_h)$.
Once this is taken into account the trion binding energy is remarkably weakly dependent on the mass ratio since
the binding energy stays approximately in the range $[0.5233 , 0.531]$ for $r$ going from $0$ to $0.7$, which
corresponds to a mass ratio $m_e/m_h$ going to from $0$ to $2.3$. This comprises most of the practical range
for standard semiconductors. This is in qualitative agreement with earlier findings \cite{sycmon}.

To a large extent this weak dependence of the trion binding energy on the mass ratio is due to the fact that the trion
is anyway weakly bound compared to the exciton binding energy itself. This is fairly clear on the $r=0$ side since,
compared to the free exciton, the additional electron is attracted by the fluctuating dipole of the exciton, which is a much
weaker attractive potential than a bare charge. Hence it is not surprising that the additional binding is roughly $6 \%$ of the 
exciton binding energy. On the $r=1$ side the hole is attracted by the two infinitely heavy electrons, and there is no longer
any quantum fluctuation in the attractive potential, so one may expect a stronger binding, as it is indeed the case.
However the resulting binding is not much stronger, as it is easily seen from the early perturbative calculations of
Morse and Stueckelberg \cite{mostu} for the equivalent hydrogen molecular ion problem, which gave a binding energy
of $0.571$ atomic units, not so far from the exact numerical result given above.

In addition to this narrow range of variation $E_b(r)$ presents a wide minimum. Indeed, as we have seen in section \ref{limca},
$E_b(r)$ decreases from its $r=0$ value, due to the 'mass-polarization correction'. This correction happens to be fairly weak
(it is zero if correlations between electrons in the ground state wave function are neglected \cite{betsal}), so $E_b(r)$ starts
from $r=0$ with a small negative slope. On the $r=1$ side $E_b(r)$ starts from a somewhat higher value, but it drops more
rapidly owing to the $E_b(1)-E_b(r) \propto (1-r)^{1/2}$ dependence we have found in section \ref{limca}. Hence it reaches
fairly rapidly values in the vicinity of the minimum. Inspection of Fig.~\ref{fig} gives the feeling that the whole curve can be
understood from these two limiting behaviours. In order to assert this statement more quantitatively we have tried to fit our
results with an analytical expression $E_b(r)=a-b\,r - c\,(1-r)^{1/2}$ involving only these behaviours. Amazingly we have found
that
\begin{eqnarray}\label{ebfit}
E_b(r)=0.71347-0.11527 \,r -0.18580\, \sqrt{1-r}
\end{eqnarray}
fits perfectly our numerical results within our estimated precision. This is plotted on Fig.~\ref{fig} as the dashed line.
Hence we can not exclude that this formula corresponds to an analytical result, although this seems extremely unlikely 
in view of the complexity of the problem. Note that it is natural to have a reasonable fit to the asymptotic behaviours on both
sides $r=0$ and $r=1$, but one would expect to need at least 4 independent constants. The surprising feature of Eq.(\ref{ebfit}) is
that it contains only 3 constants, and that moreover they provide a perfect fit over the whole range of $r$. Anyway, analytical
or not, Eq.(\ref{ebfit}) has the advantage of showing that an understanding of $E_b(r)$ in the two limiting cases $r=0$ and $r=1$
allows an understanding of the whole $E_b(r)$. This might be a useful approach for the trion problem in situations other 
than the simple 3D effective mass case that we have treated.

\section{Conclusion}

In this paper we have applied to the trion problem in semiconductors a general approach proposed previously for the three-body problem.
This has been done in the simple case of a three-dimensional semiconductor with effective mass approximation and bare Coulomb 
interaction. In this situation the trion is only weakly bound compared to the exciton. This brings practical complications to the
numerical work, but beyond these details our procedure works without any problem. Our numerical results, which extend over
the whole range of electron to hole mass ratio, are in perfect agreement with existing earlier work making use of variational methods.
In this paper we have not investigated the trion wave function, which comes out of the same numerical calculation as the
trion binding energy.  We expect this wave function to be fairly structured. We intend to proceed to this investigation in a following paper.

We have considered more specifically the cases of low or high electron-to-hole mass ratio. In these regimes the trion binding energy
has a simple analytical dependence on the mass ratio. We have made use of these simple behaviours in these limiting cases
to build an interpolation formula for the whole domain of electron-to-hole mass ratio. This formula contains only three parameters.
Very surprisingly it agrees perfectly well with our numerical results within our precision. Hence it provides in practice the complete
answer for the trion binding energy in the physical situation we have investigated. In more complicated situations this success
points to the interest of investigating limiting cases in order to understand qualitatively, and perhaps quantitatively, the trion
binding energy in the whole range of parameters of interest.


\begin{references}
\bibitem{fad}Actually we have found that our equations can be related to the work of Faddeev \cite{fadd}.
\end{references}
\end{document}